\documentclass[fleqn,usenatbib]{mnras}

\usepackage[T1]{fontenc}

\DeclareRobustCommand{\VAN}[3]{#2}
\let\VANthebibliography\thebibliography
\def\thebibliography{\DeclareRobustCommand{\VAN}[3]{##3}\VANthebibliography}

\usepackage{graphicx}	
\usepackage{amsmath}	
\usepackage{amssymb}	
\usepackage{wasysym}
\usepackage{caption}
\usepackage{subcaption}
\usepackage{hyperref}
\usepackage{iondefs}
\usepackage{threeparttable}
\usepackage{newtxtext,newtxmath}
\usepackage[]{xcolor}
\usepackage{deluxetable}
\usepackage{longtable}
\usepackage[font=small]{caption}
\newcommand{\MgIItitle}{\hbox{\bfseries Mg\kern0.1em\scshape ii}}

\definecolor{my_red}{RGB}{255, 89, 94}
\definecolor{my_purple}{RGB}{143, 120, 174}
\definecolor{my_orange}{RGB}{255, 146, 76}
\definecolor{my_green}{RGB}{138, 201, 38}
\definecolor{my_blue}{RGB}{25, 130, 196}
\definecolor{my_yellow}{RGB}{255, 202, 58}
\definecolor{my_g1_green}{RGB}{100, 167, 71}
\definecolor{my_g1_blue}{RGB}{66, 103, 172}
\definecolor{my_pink}{RGB}{199, 105, 134}

\title[Differences Between Emission and Absorption]{Differences Between Emission and Absorption Tracers of Spatially Resolved Outflows in Clumpy z$\sim$0.1 Star-forming Galaxies}

\author[A.~Fern\'{a}ndez-Figueroa et al.]
{Antonia Fern\'{a}ndez-Figueroa$^{1}$\thanks{E-mail: afernandez@das.uchile.cl},
Glenn G.~Kacprzak,$^{1}$ Deanne B. Fisher,$^{1}$ Magdalena Hamel-Bravo,$^{1}$
\newauthor{Karl Glazebrook}$^{1}$\\
$^{1}$Centre for Astrophysics and Supercomputing, Swinburne University of Technology, Hawthorn, Victoria 3122, Australia\\
}

\date{Accepted 2026 April 02. Received 2026 March 08; in original form 2025 October 26.}

\pubyear{2025}

\begin{document}
\label{firstpage}
\pagerange{\pageref{firstpage}--\pageref{lastpage}}
\maketitle

\begin{abstract}
We present spatially resolved Keck/LRIS spectroscopy of three clumpy star-forming galaxies at $z\sim0.1$, comparing outflow properties traced by {\Halpha} and {\MgII} emission with those probed by {\MgII} and {\NaID} absorption. Outflow velocities measured using {\MgII} absorption ($\langle v_{\rm out} \rangle = -560 \pm 30$~\kms) are consistently higher than those traced by {\Halpha} emission ($\langle v_{\rm out} \rangle = -124 \pm 3$~\kms) across $\sim$5~kpc$^{2}$ regions. Despite this offset, the correlation between $v_{\rm out}$ and galaxy properties, such as SFR and $\Sigma_{\rm SFR}$, show similar slopes for both tracers, with {\MgII} absorption systematically offset by $\sim 0.4$~dex. In two galaxies, {\MgII} emission is also detected, yielding velocities consistent with {\Halpha}. In one galaxy we also detect outflows in {\NaID} absorption and find similar velocities as {\MgII} in absorption, which leads to a $\sim$0.4~dex higher {\NaID} outflow velocities compared to those measured in emission. Our spatially resolved results are consistent with those found for galactic-scale measurements, implying the outflow relationships are similar from the sales of $\sim$1-2~kpc to global measurements. Combined with literature measurements, these results suggest that the offset in velocities is driven not by ionisation state, but rather by the systematics associated to how absorption and emission measures trace the gas density. 
\end{abstract}

\begin{keywords}
ISM: jets and outflows -- galaxies: evolution -- galaxies: starburst
\end{keywords}



\section{Introduction}

Star formation-driven outflows are a necessary component for galaxy evolution models to be consistent with observations of galaxies. They contribute to metal enrichment of the intergalactic medium \citep[][]{Garnett2002, Tremonti2004, Dalcanton2007}, regulate star-formation within galaxies \citep[][]{Hopkins2006, Cazzoli2014}, and play a fundamental role in shaping galaxy disks \citep[][]{Scannapieco2008}. These outflows are observed in high star-formation rate (SFR) galaxies at all redshifts \citep[][]{Veilleux2005, Forster-Schreiber2019, Carniani2024, Thompson&Heckman2024}. Given the important role outflows play in regulating star formation, it is important to understand their properties.

There are two widely used tracers of outflows: optical emission lines and UV/near-UV absorption lines. Outflows in emission lines, such as the {\OIIedblt}, {\OIIIedblt}, {\NIIedblt}, {\SIIedblt} doublets, and the Balmer series, are identified by finding broad blue shifted components in their line profiles. Because the intensity of emission line gas scales with density as $I \propto n^2$, these measurements are naturally biased towards higher-density regions of the outflow and therefore preferentially trace gas located closer to the galaxy. In contrast, absorption lines, such as {\MgIIdblt}, trace the gas linearly with density, giving access to more diffuse material, a wider range of column densities, and gas located further away along the line of sight \citep[][]{Veilleux2005}. Measurements of outflows in absorption are often obtained by using the galaxy itself as a background source, a technique known as `down-the-barrel' spectroscopy \citep[e.g., ][]{Martin2005, Martin&Bouche2009, Bordoloi_2011a, Kornei2012, Kacprzak2014, Rubin_2014, Heckman2015}. Direct comparison of emission and absorption tracers can therefore not only reveal the biases inherent to each method, but also provide a more complete view of the outflow.

The vast majority of resolved outflow measurements are made on emission lines. Work on resolved observations of outflows have recently shown strong correlations between outflow velocity ($v_{\rm out}$) and star-formation rate surface density \citep[$\Sigma_{\rm SFR}$, ][]{ReichardtChu2022, ReichardtChu2025}, which was unclear in unresolved studies \citep[][]{Rupke2005, Chen2010, Kornei2012, Newman2012, Arribas2014, Bordoloi2014, Rubin_2014, Chisholm2015, Heckman2015, Fernandez-Figueroa2025}. These correlations are interpreted to favour models of feedback and test simulations \citep[e.g., ][]{Kim2020, Rathjen2023}. However, it is unclear whether these resolved correlations remain for absorption lines, or if they are weakened as this gas tracer favours larger distances from the galaxy.

To date, only two resolved studies comparing outflow tracers have been performed in the literature \citep[][]{Wood2015, MartinFernandez2016}. Both works focus on singular galaxies and find that outflow velocities traced in emission are systematically lower than those traced in absorption. More recently, \citet{Xu2025} and \citet{Peng2025} performed similar analyses for unresolved spectra of galaxies, finding that while $v_{\rm out}$ scales with SFR for both tracers, absorption consistently yields higher velocities than emission.

In this Letter, we use spatially resolved observations of three clumpy galaxies at $z \sim 0.1$ to investigate the relationship between outflow properties traced by absorption to those measured in emission. We assume a flat $\Lambda$ cold dark matter ($\Lambda$CDM) Universe with $\Omega_{\Lambda}= 0.7$, $\Omega_{M} = 0.3$, and $H_{0} = 70$~km~s$^{-1}$~Mpc$^{-1}$, and a Kroupa initial mass function \citep{Kroupa_2001b}.

\begin{table*}
	\centering
	\caption{Galaxy clump properties}
	\label{tab:values}
	\begin{tabular}{lccccccc}
      \hline
      Galaxy & $v_{\rm out,\Halpha}$ & $v_{\rm out,\MgII,abs}$ & $v_{\rm out,\MgII,em}$ & $v_{\rm out,\NaI{\rm D}}$ & SFR & $\Sigma_{\rm SFR}$\\[2pt]
      & (\kms) & (\kms) & (\kms) & (\kms) & ($M_{\odot}~{\rm yr}^{-1}$) & ($M_{\odot}~{\rm yr}^{-1}~{\rm kpc}^{-2}$)\\
      & (1) & (2) & (3) & (4) & (5) & (6) \\
      \hline
      H10-2 & $-243\pm11$ & $-523\pm174$& $-307\pm61$ & - & $0.29$ & $0.048$\\[2pt]
       & $-220\pm8$ & $-732\pm133$& $-287\pm52$ & - & $0.97$ & $0.161$\\[2pt]
       & $-221\pm8$ & $-677\pm123$& $-307\pm60$ & - & $0.96$ & $0.16$\\[2pt]
       & $-242\pm27$ & $-711\pm30$& $-326\pm27$ & - & $1.31$ & $0.218$\\[2pt]
       & $-281\pm16$ & $-685\pm76$& $-433\pm40$ & - & $0.94$ & $0.156$\\[2pt]
       & $-239\pm11$ & $-855\pm332$& $-466\pm91$ & - & $0.52$ & $0.086$\\[2pt]
       & $-349\pm32$ & $-727\pm213$& $-599\pm83$ & - & $0.03$ & $0.005$\\[2pt]
      \hline
      G13-1 & $-363\pm2$ & $-563\pm20$& $-281\pm21$ & - & $1.14$ & $0.217$\\[2pt]
       & $-445\pm1$ & $-551\pm18$& $-353\pm17$ & - & $1.13$ & $0.215$\\[2pt]
       & $-350\pm2$ & $-532\pm26$& $-253\pm17$ & - & $0.47$ & $0.089$\\[2pt]
       & $-327\pm3$ & $-619\pm21$& $-306\pm58$ & - & $0.41$ & $0.078$\\[2pt]
       & $-213\pm1$ & $-790\pm60$& $-122\pm31$ & - & $0.36$ & $0.068$\\[2pt]
      \hline
      G14-1 & $-340\pm2$ & $-632\pm20$& - & $-507\pm51$ & $0.9$ & $0.189$\\[2pt]
       & $-333\pm1$ & $-707\pm17$& - & $-598\pm45$ & $1.78$ & $0.374$\\[2pt]
       & $-312\pm3$ & $-665\pm28$& - & $-486\pm120$ & $0.93$ & $0.195$\\[2pt]
       & $-193\pm1$ & $-457\pm18$& - & $-501\pm81$ & $0.34$ & $0.071$\\[2pt]
      \hline
\end{tabular}

\raggedright The columns are: (1) Galaxy name, (2) outflow velocity traced by {\Halpha} emission, (3) outflow velocity traced by {\MgII} absorption, (4) outflow velocity traced by {\MgII} emission, (5) outflow velocity traced by {\NaI~D} absorption, (6) star-formation rate, and (7) star-formation rate surface density.
\end{table*}

\begin{figure}
     \centering
     \includegraphics[width=\columnwidth]{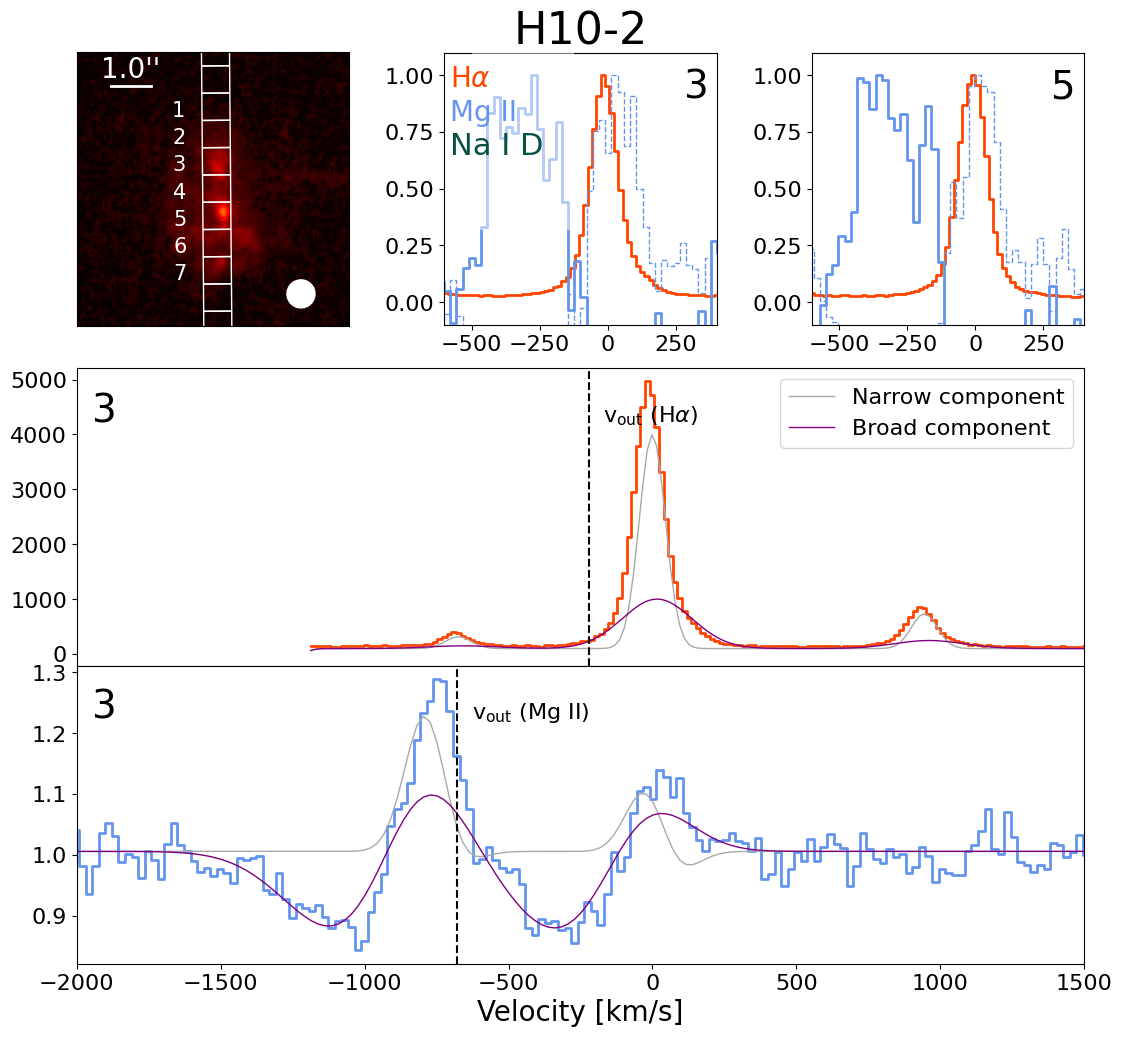}
     \caption{\textit{Top left:} \textit{HST} narrowband imaging of galaxy H10-2. The position of our LRIS slit is displayed as white lines. The white circle represents the seeing during the night of observations. \textit{Top middle and right:} Comparison of {\Halpha} emission (red), {\MgII}$\lambda2803$ absorption (solid blue) and {\MgII}$\lambda2803$ emission (dashed blue). Each panel represents a seeing width (0.68'' or 2.02~kpc) along the LRIS in selected pixels. The number in the top right corner corresponds to the pixel number. The {\Halpha} and {\MgII} have been normalised for a more direct comparison, with the {\MgII} absorption profile being inverted. The displayed profiles include both the narrow and broad components from our spectral line modelling. For all panels, the zero-velocity point is defined as the central velocity of the {\Halpha} narrow component at the corresponding pixel. We find that {\MgII} extends to much larger velocities than {\Halpha}. \textit{Middle:} {\Halpha} and {\NII} emission lines in a selected pixel, displayed in red. The narrow and broad emission components from our emission line modelling are shown in grey and purple, respectively. The measured {\Halpha} outflow velocity in this pixel is indicated with a black vertical line. \textit{Bottom:} {\MgII} emission and absorption profiles in a selected pixel, shown in blue. The narrow and broad emission components from our emission line modelling are shown in grey and purple, respectively. The measured {\MgII} absorption outflow velocity in this pixel is indicated with a black vertical line.}
     \label{fig:lines_H10-2}
\end{figure}

\begin{figure}
     \centering
     \includegraphics[width=\columnwidth]{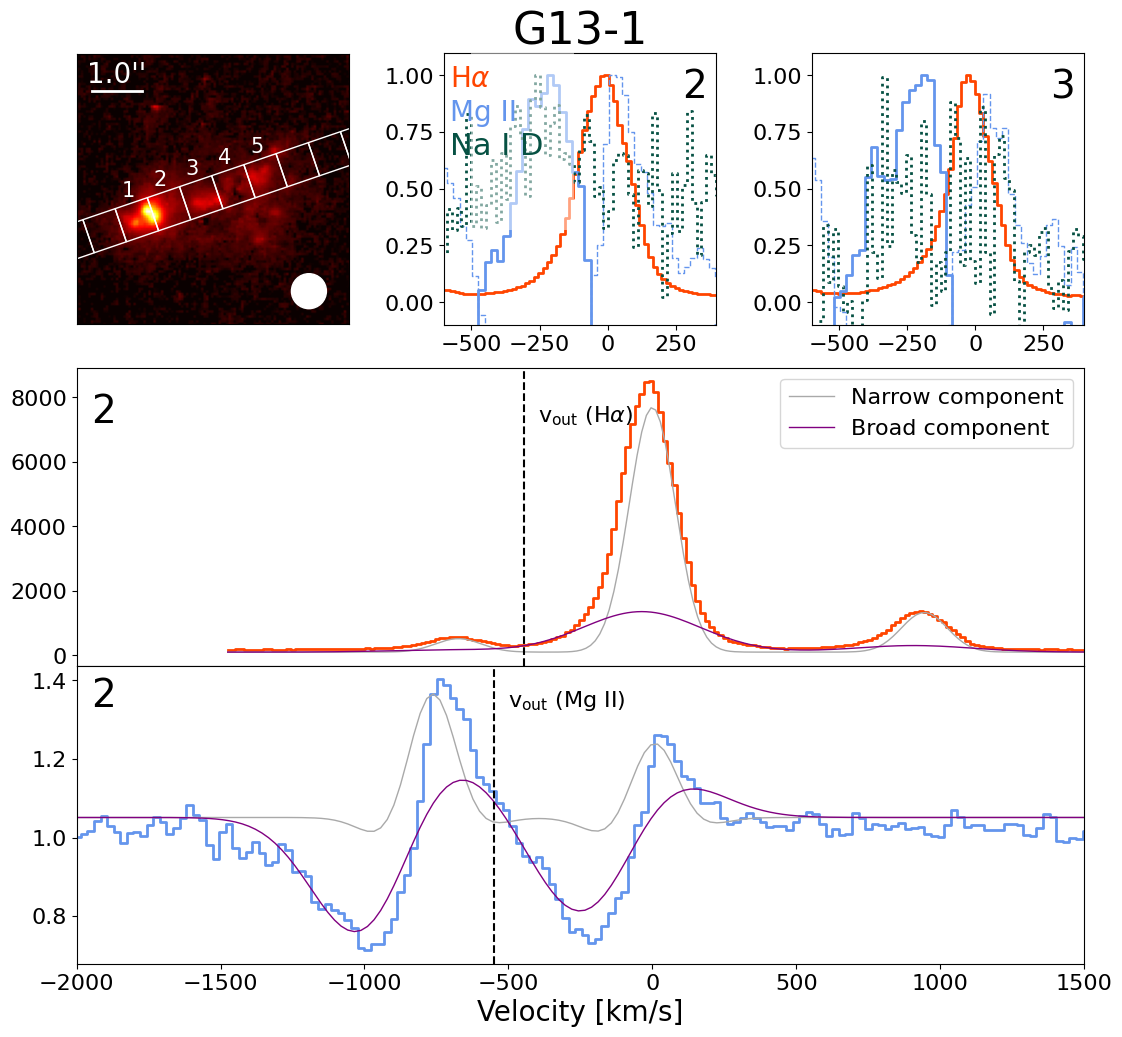}
     \caption{Same as Fig.~\ref{fig:lines_H10-2}, except for galaxy G13-1. The five spatial pixels are 0.68'' or 1.89~kpc in length. Again, we find {\MgII} absorption extending to larger velocities.}
     \label{fig:lines_G13-1}
\end{figure}

\begin{figure}
     \centering
     \includegraphics[width=\columnwidth]{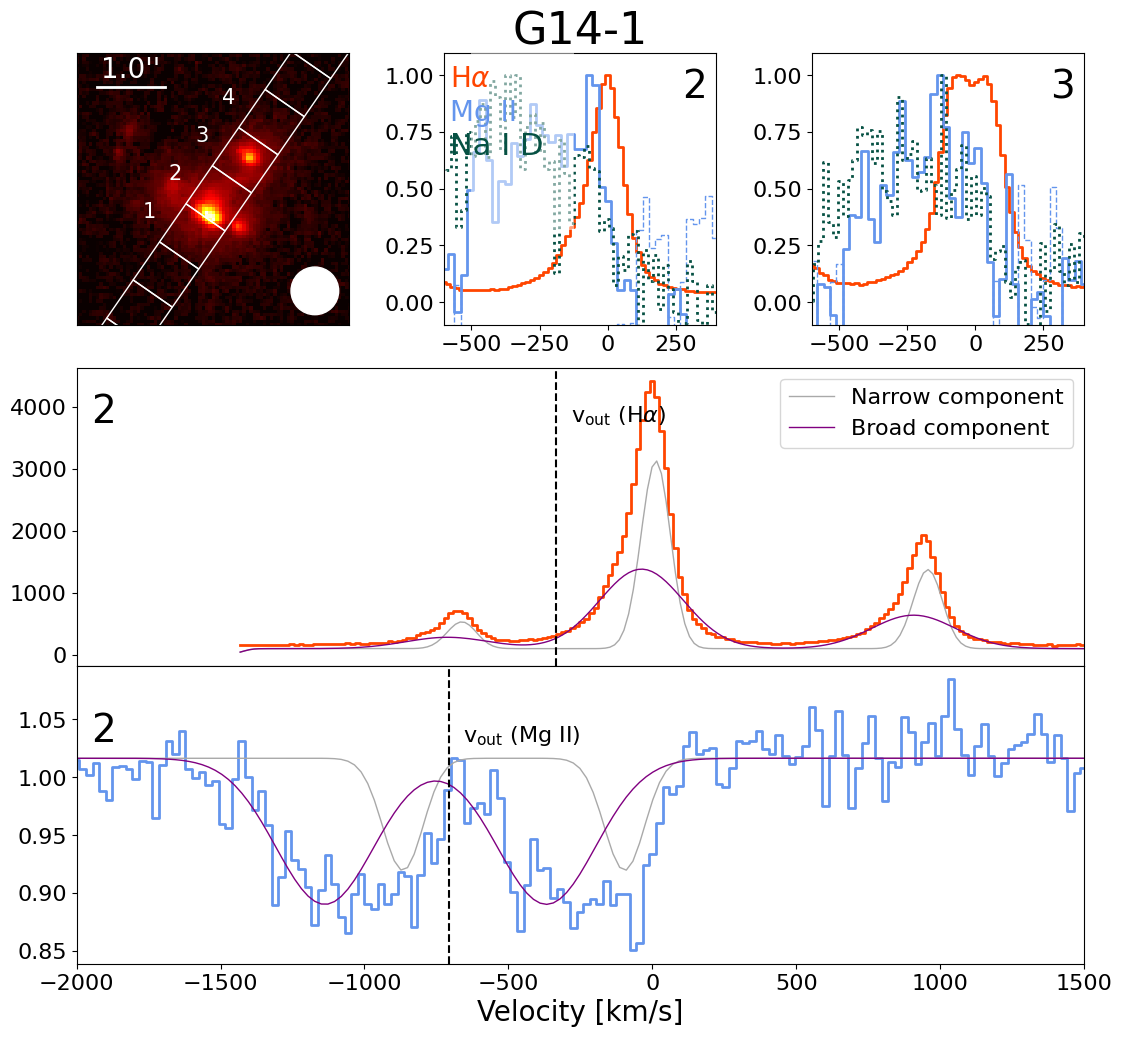}
     \caption{Same as Fig.~\ref{fig:lines_H10-2}, except for galaxy G14-1. The four spatial pixels are 068'' or 1.79~kpc in length. As in previous cases, {\MgII} reaches larger velocities. We detect {\NaI~D} absorption in this galaxy. This absorption profile is inverted and displayed in dark green. Interestingly, the velocities traced by {\NaI~D} trace very closely with those traced by {\MgII} absorption. The upturn seen in the {\Halpha} profile near $-500$~{\kms} corresponds to the {\NII} emission line.}
     \label{fig:lines_G14-1}
\end{figure}

\section{Data} \label{data}


The galaxies in the present work were selected from the DYnamic of Newly-Assembled Massive Objects \citep[DYNAMO, ][]{Green2014} survey. The DYNAMO survey consists of $95$ star-forming galaxies at $z\sim0.1$, covering a wide range of SFRs ($0.02-100~M_{\odot}~{\rm yr}^{-1}$) and stellar masses ($10^{9} - 10^{11} M_{\odot}$). For our study, we selected three of these galaxies: H10-2 ($z=0.1491$, $\log (M / M_{\odot}) = 9.98$, $\rm SFR = 25.4~M_{\odot}$~yr$^{-1}$), G13-1 ($z=0.1388$, $\log (M / M_{\odot}) = 10.05$, $\rm SFR = 26.5~M_{\odot}$~yr$^{-1}$), and G14-1 ($z=0.1323$, $\log (M / M_{\odot}) = 10.35$, $\rm SFR = 8.3~M_{\odot}$~yr$^{-1}$). These targets were selected due to their clumpy morphology, and high SFRs, making them appropiate to study star formation-driven outflows. They were also selected to be at a redshift $z \geq 0.1$, where {\MgII} is feasibly observable with Keck/LRIS. 

The galaxies were observed on 28 March 2017 (PID: 2017A\_W246) using Low Resolution Imaging Spectrometer \citep[LRIS; ][]{OKE1995} on the Keck telescope with the 0.7$''$ slit. We employed the B1200/3400 grism in the blue CCD, resulting in a wavelength range of 2850-3852~{\AA} and a spectral resolution of 50 {\kms}. Simultaneously, we used the R1200/3400 grating with a central wavelength of $7003$~{\AA} in the red CCD, resulting in a wavelength range of 6190-7840~{\AA}. These wavelength ranges cover the {\MgIIdblt} absorption and {\Halpha} emission lines. The mean galaxy exposure times were 6177~s for the blue side and 5445~s for the red side. The spectra were reduced using standard IRAF routines, and were heliocentric and vacuum corrected. The spectra were then spatially binned to match the seeing on the night of observations. The location of the LRIS slits, along with examples of {\Halpha} emission and {\MgII} absorption are shown in Figs.~\ref{fig:lines_H10-2},~\ref{fig:lines_G13-1} and~\ref{fig:lines_G14-1}.

These observations are complemented by \textit{HST} imaging taken with the Wide Field Camera on the Advanced Camera for Surveys (WFC/ACS). The data were obtained using the ramp filters FR716N and FR782N, which were centred in the {\Halpha} emission lines of our galaxies. Each target was observed for 45~min. More details on the imaging acquisition and reduction can be found in \citet{Fisher2017}.

We converted the counts in our LRIS data into physical flux units by matching them with the \textit{HST} narrowband imaging, following the method detailed in \citet{HamelBravo2025}. The \textit{HST} filter was first applied to the LRIS spectra to simulate the corresponding \textit{HST} observations, from which we generate light profiles along the slit. We then determined the slit position on the \textit{HST} image by creating simulated light profiles using a synthetic slit with the same dimensions and position angle, varying its location until the best match with the observed LRIS profile was found, ensuring that the brightest part of the galaxy is included. Finally, we compute the scaling factor between the LRIS and \textit{HST} profiles and apply it along the spatial axis of the LRIS data to obtain flux-calibrated {\Halpha} emission lines.

Given the relationship between resolved SFR and outflow velocity \citep[][]{ReichardtChu2025}, we study how this relationship compares for outflows measured by different tracers. The SFR on each pixel was calculated by measuring the {\Halpha} emission flux and applying the equation from \citet{Hao_2011b}: ${\rm SFR} = C_{\tiny \Halpha} L_{\tiny \Halpha} 10^{-0.4 A_{\tiny \Halpha}}$, where $L_{\tiny \Halpha}$ is the observed {\Halpha} luminosity, $C_{\tiny \Halpha}=10^{-41.257}$, and $A_{\tiny \Halpha}$ is the {\Halpha} extinction parameter of each galaxy, adopted from \citet{Fisher2017}. The calculated values along each seeing bin are presented in Table~\ref{tab:values}.

\begin{figure}
     \centering
     \includegraphics[width=\columnwidth]{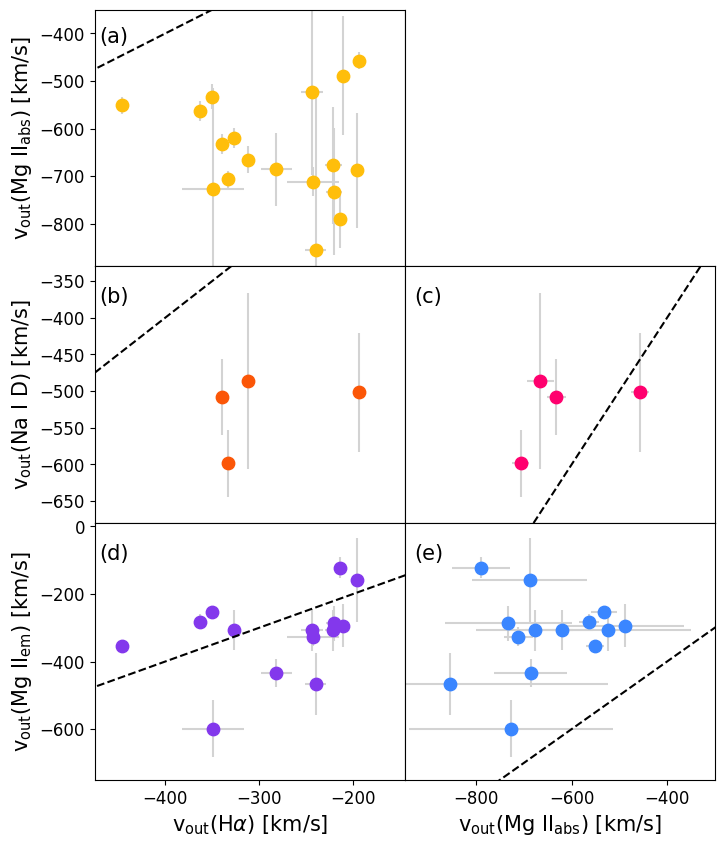}
     \caption{Comparison of outflow velocities measured by different tracers. The panels show: \textit{(a)} {\MgII} absorption vs. {\Halpha} emission, \textit{(b)} {\NaID} absorption vs. {\Halpha} emission, \textit{(c)} {\NaID} absorption vs. {\MgII} absorption, \textit{(d)} {\MgII} emission vs. {\Halpha} emission, and \textit{(e)} {\MgII} emission vs. {\MgII} absorption. The black dashed lines represent the 1:1 relationship. We find that absorption measured outflows have consistently higher outflow velocities. Emission line tracers of outflows show similar velocities to each other, and absorption line tracers yield consistent velocities to each other.}
     \label{fig:corner_plots}
\end{figure}

\begin{figure}
     \begin{subfigure}[b]{\columnwidth}
     \centering
         \includegraphics[width=\columnwidth]{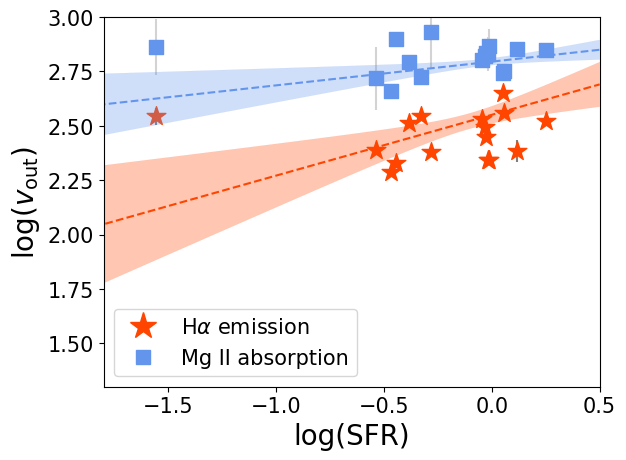}
     \end{subfigure}
     
     \begin{subfigure}[b]{\columnwidth}
         \centering
         \includegraphics[width=\columnwidth]{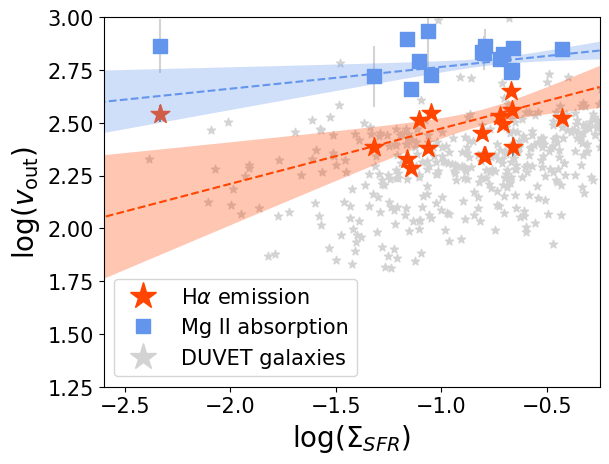}
     \end{subfigure}

     \caption{\textit{Top:} Comparison of outflow velocities with SFR. Outflows traced by {\Halpha} emission are represented as red stars, and those traced by {\MgII} absorption are displayed as blue squares. Linear fits to each sample are shown as red and blue dashed lines. \textit{Bottom:} Comparison of outflow velocities with $\Sigma_{\rm SFR}$. Measurements of DUVET galaxies \citep[][]{ReichardtChu2025} are displayed as light grey stars.}
     \label{fig:scalling_rel}
\end{figure}

\section{Spectral line modeling} \label{modelling}

We modelled the {\Halpha} and {\NIIedblt} emission lines by two sets of three Gaussians to the spectral profiles, with each set representing an ISM narrow component and a broad component associated with outflowing material. We measure typical velocity dispersions of $50$~{\kms} for the narrow component, which is approximately in the range of previous DYNAMO measurements \citep[][]{Green2014}. The narrow and broad {\Halpha} components were modelled independently, with amplitude, velocity and velocity dispersion free to vary. The {\NII} components were set to have the same velocities and velocity dispersions as the {\Halpha} lines. For each spectrum, we ran 1000 Monte Carlo realisations, allowing the data to vary within a Gaussian distribution with a standard deviation equal to the $1\sigma$ errors of the spectra. We adopted the median and standard deviation of the resulting distributions as the final parameters values and uncertainties.

Similarly, we modelled the {\MgIIdblt} absorption and emission lines by fitting narrow Gaussians to represent the ISM and stellar absorption, along with broad Gaussians associated with outflows. The narrow absorption and emission components were modelled based on the narrow {\Halpha} component, allowed to vary within a small offset in velocity and velocity dispersion. The broad components were allowed to vary more freely, but still using the corresponding {\Halpha} values as a starting guess. The two lines of the doublet were constrained to have the same velocities and velocity dispersions, but their amplitudes were fitted independently. It is worth noting that the presence of {\MgII} in both absorption and emission simultaneously leads to large systematic uncertainties in our modelling. In the case of G14-1, where no {\MgII} emission was detected, only the absorption was fitted. Examples of the spectral line modelling are displayed in the middle and bottom panels of Figs.~\ref{fig:lines_H10-2},~\ref{fig:lines_G13-1} and~\ref{fig:lines_G14-1}.

We corrected for beam smearing while modelling these spectral lines. The kinematic properties of G14-1, G13-1 and H10-2 have been obtained by modelling the kinematics using an arctan model \citep{Green2014, Bekiaris2016, Girard2021}. We used the kinematic 2D models to obtain the unresolved velocity shear, or contribution from the beam-smearing, in each spaxel. We combine the shear values (or beam-smearing contributions) to the instrumental broadening effect and then convolved directly the gaussian models to the contribution of these two effects when fitting the spectral lines of our observations. This correction affects mostly the central 1--2 resolution elements, where the velocity gradient is at its steepest. This is a similar approach to the beam-smearing and instrumental broadening corrections applied in previous studies \citep[][]{Epinat2012, Bassett2014, Green2014, Levy2018}, but takes into account the asymmetric spectral line shape that the beam-smearing can cause.


For our broad outflow components, we define $v_{\rm out} = v - 2 \sigma$, where $v$ is the central velocity of the component and $\sigma$ is the velocity dispersion. This definition captures the motion of the bulk of the outflowing gas, and has been previously used in the literature by similar works \citep[e.g., ][]{Genzel2011, Davies2019, ReichardtChu2022}. Uncertainties are derived from our previously described Monte Carlo realisations and propagated accordingly. These values are presented in Table~\ref{tab:values} and Fig.~\ref{fig:corner_plots}, comparing $v_{\rm out}$ measured in different tracers. 

\section{Results}

In Figs.~\ref{fig:lines_H10-2}, \ref{fig:lines_G13-1} and \ref{fig:lines_G14-1} we display a visual comparison of {\Halpha} emission and {\MgII} absorption velocity profiles, shown in red and blue, respectively. It is clear from the figures that the majority of the {\MgII} absorption lies at higher velocities than the {\Halpha} emission. Moreover, Fig.~\ref{fig:lines_G14-1} displays detected {\NaID} absorption velocities, displayed in green. These velocities closely match those found in {\MgII} absorption. 
It is worth noting that the displayed profiles are not corrected for the rotation of the galaxies themselves, but this effect is accounted for in our outflow modelling.

The results from our line modelling described in Section~\ref{modelling} are presented in Table~\ref{tab:values} and panel (a) of Fig.~\ref{fig:corner_plots}. The mean $v_{\rm out}$ traced by {\Halpha} emission is $v_{\rm out} = -256\pm7$, $-254\pm1$, and $-295\pm1$~{\kms} for H10-2, G13-1 and G14-1, respectively. In parallel, we find that the mean $v_{\rm out}$ traced by {\MgII} absorption is $v_{\rm out} = -701\pm68$, $-611\pm15$, and $-615\pm11$~{\kms} for the same galaxies. These values are higher than those measured in the literature for unresolved galaxies with similar star-formation rates as those in our sample \citep[$\sim$400 \kms, ][]{Rubin_2014}. Unlike the \citet{Rubin_2014} sample, the DYNAMO galaxies are selected by clumpiness and high velocity dispersion. These may correlate with higher outflow velocities. Our comparison of $v_{\rm out}$ measured in {\MgII} absorption and {\Halpha} emission yields differences by factors of $2.7\pm0.3$, $2.4\pm0.1$, and $2.1\pm0.1$. Similar trends have been observed in previous works: \citet{ReichardtChu2022} measured outflow velocities of $\sim250-300$~{\kms} in the centre of IRAS08339+6517 using {\OIIIe} and {\Hbeta} emission lines, while \citet{Chisholm2015} measures a $v_{90}$ in absorption lines from the same galaxy at $\sim1000$~{\kms}. This results in a difference by a factor of 3.4, which is consistent with our findings. 

In addition to absorption, we detect {\MgII} in emission in two of our galaxies. The mean outflow velocities are $v_{\rm out} = -389\pm24$ {\kms} for H10-2 and $v_{\rm out} = -263\pm15$ {\kms} for G13-1, as listed in Table~\ref{tab:values}. Panel (d) of Fig.~\ref{fig:corner_plots} compares $v_{\rm out}$ measured in {\Halpha} emission with that measured in {\MgII} emission. We find that these measurements are positively correlated and have similar magnitudes, indicating that {\MgII} emission traces a kinematically similar, and likely closely related, outflow component to that traced by {\Halpha}. Given that Mg II is a resonant transition, its emission may be affected by scattering, so this agreement does not necessarily imply that the two lines arise from the same gas parcels. Moreover, panel (e) of Fig.~\ref{fig:corner_plots} shows a comparison between $v_{\rm out}$ measured in {\MgII} emission and absorption. Once again, the measurements performed in absorption are consistently higher than those traced by emission. However, it is important to note that there is a significant degeneracy between the {\MgII} absorption and emission components, so these results must be interpreted with caution.


We detect {\NaID} absorption in G14-1. As presented in Table~\ref{tab:values}, its mean outflow velocity is $v_{\rm out} = -523\pm 40$ {\kms}. In panel (c) of Fig.~\ref{fig:corner_plots}, we compare these outflow velocities with those measured in {\MgII} absorption. We find that these velocities are consistent with each other. Additionally, the panel (b) of Fig.~\ref{fig:corner_plots} compares outflow velocities measured with {\NaID} absorption with those probed by {\Halpha} emission. In this case, the outflow velocities traced by {\Halpha} emission are consistently lower than those measured in {\NaID}. These results indicate that absorption diagnostics probe similar components of the outflow.

     


We present our derived SFR values in Table~\ref{tab:values}. These values are calculated within one spatial resolution element. In the case of H10-2, SFRs range from $1.3$ to {$0.03~M_{\odot}~{\rm yr}^{-1}$}, with an average of {${\rm SFR} = 0.7~M_{\odot}~{\rm yr}^{-1}$}, calculated over an area of 6.0~kpc$^{2}$. In G13-1, SFR values span from $1.1$ to {$0.4~M_{\odot}~{\rm yr}^{-1}$}, having a mean value of {${\rm SFR} = 0.7~M_{\odot}~{\rm yr}^{-1}$}, measured across 5.3~kpc$^{2}$. For G14-1, we measure SFRs between $0.3$ and {$1.8~M_{\odot}~{\rm yr}^{-1}$}, averaging a {${\rm SFR} = 1~M_{\odot}~{\rm yr}^{-1}$}, within 4.8~kpc$^{2}$.

In the top panel Fig.~\ref{fig:scalling_rel}, we display $v_{\rm out}$ as a function of SFR. $v_{\rm out}$ traced by {\MgII} absorption and {\Halpha} emission are shown as squares and stars, respectively. For a fixed SFR, $v_{\rm out}$ traced by {\MgII} absorption is consistently higher than that traced by {\Halpha} emission, which is in agreement with previous results found in the literature \citep[][]{Wood2015, Xu2025}. Moreover, $v_{\rm out}$ increases with SFR for both tracers, which is consistent with previous results performed in unresolved galaxies \citep[][]{Rupke2005, Chen2010, Newman2012, Bordoloi2014, Rubin_2014, Chisholm2015}. We perform linear fits of the form: {$\log v_{\rm out} = a \log {\rm SFR} + b$}, which are shown in the figure as red and blue dashed lines. The best fit slope for outflows measured in emission is {$a_{\Halpha} = 0.28 \pm 0.22$}, while that of outflows traced by absorption is {$a_{\MgII} = 0.11 \pm 0.08$}. These slopes are consistent within errors, although the scatter is large.

While SFR is an important quantity, $\Sigma_{\rm SFR}$ provides information about the distribution of energy capable of driving outflows. Galaxies with $\Sigma_{\rm SFR} \geq 0.1$ are more likely to be undergoing outflows \citep[][]{Heckman2002, Heckman2015, ReichardtChu2022}. Given the SFR and effective radius of our galaxies \citep[][]{Fisher2017}, we expect all of them to be above this threshold. We calculate $\Sigma_{\rm SFR}$ by computing the area in kpc enclosed on each pixel, and dividing the SFR by this quantity. As presented in Table~\ref{tab:values}, all of the galaxies in our sample have regions with $\Sigma_{\rm SFR}$ above this threshold. H10-2 spans from {$\Sigma_{\rm SFR} = 0.01$} to {$0.2~M_{\odot}~{\rm yr}^{-1}~{\rm kpc}^{2}$}. In the case of G13-1, values range form {$\Sigma_{\rm SFR} = 0.1$} to {$0.2~M_{\odot}~{\rm yr}^{-1}~{\rm kpc}^{2}$}. For G14-1, we measure from {$\Sigma_{\rm SFR} = 0.1$} to {$0.4~M_{\odot}~{\rm yr}^{-1}~{\rm kpc}^{2}$}.

The bottom panel of Fig.~\ref{fig:scalling_rel} shows $v_{\rm out}$ as a function of $\Sigma_{\rm SFR}$ for both of our diagnostics. As it is expected from the literature, $v_{\rm out}$ increases with $\Sigma_{\rm SFR}$ for both tracers. We model the $v_{\rm out}$ versus $\Sigma_{\rm SFR}$ relationship using a linear model of the form: {$\log v_{\rm out} = a \log \Sigma_{\rm SFR} + b$}, presented as red and blue dashed lines in the figure. For outflows in emission, the best fit slope is $a_{\Halpha} = 0.26 \pm 0.16$, compared to $a_{\MgII} = 0.10 \pm 0.08$ for absorption. Once again, these slopes are consistent within errors, but the scatter is large.

\citet{ReichardtChu2025} present a resolved study of outflows in ten starburst galaxies from the DUVET  galaxy survey, using {\Hbeta} and {\OIIIe$\lambda5007$} emission lines. Their results are shown in Fig.~\ref{fig:scalling_rel} as grey stars. They report a shallow correlation between $v_{\rm out}$ and $\Sigma_{\rm SFR}$ with {$v_{\rm out}(\Hbeta) \propto \Sigma_{\rm SFR}^{0.2}$}. We compare these results by running a 2D KS test between their sample and our {\Halpha} measurements, which yields a p-value of $10^{4}$, suggesting that both samples are unlikely to be drawn from the same underlying distribution.

\section{Discussion and conclusions} \label{conclusions}

In this Letter we study outflows on three clumpy star-forming galaxies at $z\sim0.1$ selected from the DYNAMO survey. Using spatially resolved observations, we measure outflow velocities ($v_{\rm out}$) employing two tracers: {\Halpha} emission and {\MgII} absorption. We find that outflow velocities traced by absorption are consistently higher than those traced by emission.


Only two previous works in the literature have performed resolved studies of outflows comparing emission and absorption features. \citet{Wood2015} measured $v_{\rm out} = -290$~{\kms} in {\Halpha} emission and $v_{\rm out} = -900$~{\kms} in absorption, concluding that a large fraction of the outflowing gas is too diffuse to be detected in emission. They also reported evidence that the line width in the {\Halpha} emission originates from shocks near the disc of the galaxy, suggesting that {\Halpha} might not be an ideal tracer of outflows. Similarly, \citet{MartinFernandez2016} studied a single galaxy and found $v_{\rm out} = -300$~{\kms} in {\Halpha} emission and $v_{\rm out} = -470$~{\kms} in absorption. They also found that the outflow detected in emission is compact and concentrated in a region near the galactic centre, consistent with emission tracing a denser, lower-velocity gas. In contrast, more extended, diffuse gas remains undetected in emission. Our results expand on these studies by increasing the sample size by 150\%, with resolved outflows in three galaxies. We measure $v_{\rm out} = -291$~{\kms} in {\Halpha} emission and $v_{\rm out} = -651$~{\kms} in absorption. This reinforces the conclusion that different tracers probe different phases of the outflow, with absorption being sensitive to a more diffuse, higher velocity component.


Two recent studies have compared outflow velocities traced in emission and absorption for spatially unresolved spectra of galaxies. \citet{Xu2025} studied 33 galaxies and found a correlation between the two tracers: {$v_{\rm out}(\Halpha) = 0.68 v_{\rm out}(\rm abs)$}. They also reported that both tracers scale with SFR, but velocities traced by emission are $\sim 0.2$~dex lower. They suggest that discrepancies in previous high-$z$ studies might be due to them being performed on emission lines, which are less sensitive to diffuse gas. In parallel, \citet{Peng2025} analysed 15 dwarf galaxies and found a correlation in the velocities measured by the two tracers: {$v_{\rm out}(\Halpha) = 0.8 v_{\rm out}(\rm abs) + 0.3$}. Interestingly, both studies use a variety of absorption lines with a wide range of ionisation potentials, suggesting that these differences are associated with the applied methods and not with the gas density or temperature. Our results are consistent with these findings, indicating that these are both local and global trends. It is worth noting that the non-resolved studies do not take into account the rotation of the galaxies themselves when modelling their emission and absorption lines, whereas we do.


There is a general assumption that emission line gas traces denser gas and is thus probing a component of the outflow that is located in closer proximity to the launch site. The offset in velocities that we observe, consistent with other works, suggests that this interpretation is valid. This is further supported by the fact that {\Halpha} and {\MgII} emission trace similar velocities (although {\MgII} emission may be modified by resonance scattering), while {\MgII} and {\NaID} absorption do as well, indicating that the observed difference does not depend on density or ionisation potential. Moreover, there is the same offset in other studies making this comparison, which use a variety of absorption line tracers. Though we do not know the timescales associated with the gas, it is plausible to expect higher velocity gas to travel farther from the galaxy. Absorption-emission line comparisons of outflows, therefore, offer an interesting and more holistic view of the outflow for comparison to simulations. Future work is necessary to extend this analysis to higher redshifts and test whether these trends persist.


\section*{Acknowledgements}

Some of the data presented herein were obtained at the W. M. Keck Observatory, which is operated as a scientific partnership among the California Institute of Technology, the University of California, and the National Aeronautics and Space Administration. The Observatory was made possible by the generous financial support of the W. M. Keck Foundation. Observations were supported by Swinburne Keck program 2017A\_W246. The authors wish to recognize and acknowledge the very significant cultural role and reverence that the summit of Maunakea has always had within the indigenous Hawaiian community. We are most fortunate to have the opportunity to conduct observations from this mountain.


\section*{Data Availability}

The data used in this paper will be shared following mutually agreeable arrangements with the corresponding authors.
 



\bibliographystyle{mnras}
\bibliography{refs}








\bsp	
\label{lastpage}
\end{document}